\documentclass[twocolumn,superscriptaddress,letter]{revtex4}

\usepackage{graphicx}
\usepackage{url}

\usepackage{times}
\usepackage{helvet}
\usepackage{courier}
\usepackage{graphicx}
\usepackage{amsmath}
\usepackage{amssymb}
\begin{document}

\title{Stealing Reality}

\author{Yaniv Altshuler}
\affiliation{Deutsche Telekom Laboratories, Ben Gurion University, Beer Sheva 84105, Israel}
\author{Nadav Aharony}
\affiliation{The Media Laboratory, Massachusetts Institute of Technology, Cambridge, MA 02139, USA}%
\author{Yuval Elovici}
\affiliation{Deutsche Telekom Laboratories, Ben Gurion University, Beer Sheva 84105, Israel}
\author{Alex Pentland}
\affiliation{The Media Laboratory, Massachusetts Institute of Technology, Cambridge, MA 02139, USA}%
\author{Manuel Cebrian}
\affiliation{The Media Laboratory, Massachusetts Institute of Technology, Cambridge, MA 02139, USA}%

\begin{abstract}
In this paper we discuss the threat of malware targeted at extracting information
about the relationships in a real-world social network as well as characteristic
information about the individuals in the network, which we dub \emph{Stealing
Reality}. We present Stealing Reality (SR), explain why it differs from
traditional types of network attacks, and discuss why its impact is significantly
more dangerous than that of other attacks. We also present our initial analysis
and results regarding the form that an SR attack might take, with the goal of
promoting the discussion of defending against such an attack, or even just
detecting the fact that one has already occurred.
\end{abstract}

\maketitle

\begin{figure*}[htb]
   \centering \includegraphics[width=\textwidth]{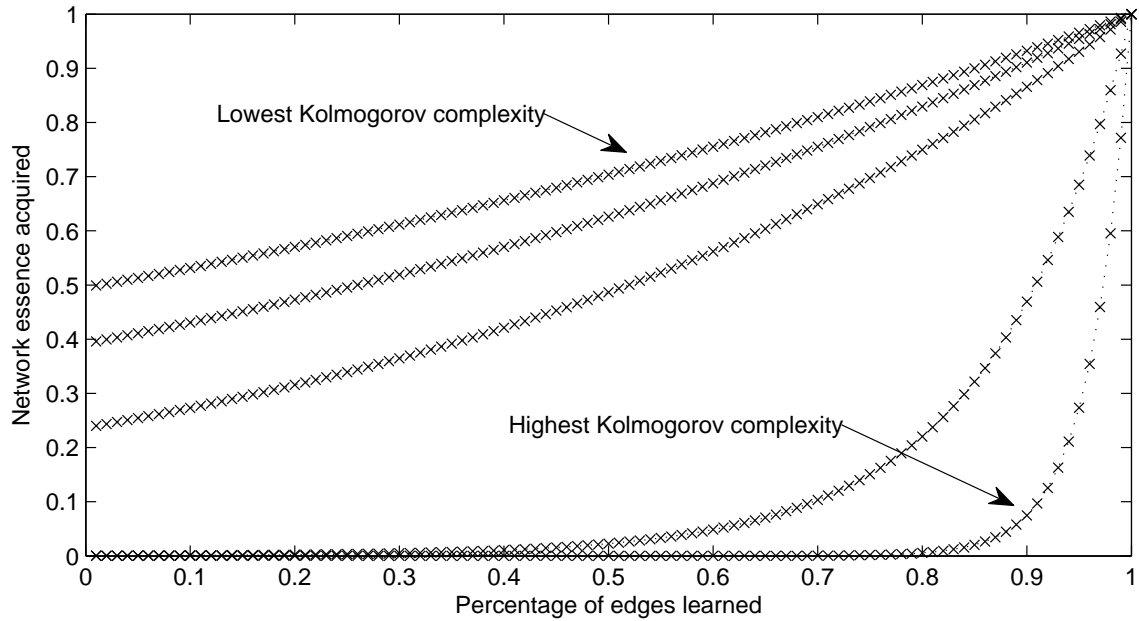}
   \caption{The evolution of $\Lambda_{S}$ as a function of the overall
   percentage of edges learned, for networks of same number of edges, but
   different values of Kolmogorov complexity. }
   \label{fig.res2}
\end{figure*}

\section{Introduction}\label{sec.intro}

History has shown that whenever something has a tangible value associated with
it, there will always be those who try to malevolently `game' the system for
profit. These days, the field of social networks is experiencing exponential
growth in popularity while in parallel, computational social science
\cite{CSS-Lazer-Science-2009} and network science
\cite{CSS-BarabasiAlbert-Science-1999,CSS-Watts-Nature-1998,CSS-Newman-SIAM-2003}
are providing real-world applicable methods and results with a demonstrated
monetary value. We conjecture that the world will increasingly see malware
integrating tools and mechanism from network science into its arsenal, as well as
attacks that directly target human-network information as a goal rather than a
means. Paraphrasing Marshall McLuhan's ``the medium is the message,'' we have
reached the stage where, now, ``the network is the message.''

Social networking concepts could be discussed both in the context of malware's
means of spreading, as well as in the context of its target goal. Many existing
viruses and worms use primitive forms of `social engineering' \cite{Granger-2001}
 as a means of spreading, in order to gain the trust of their next victims and
cause them to click on a link or install an application. For example, `Happy99'
was one of the first viruses to attach itself to outgoing emails, thus increasing
the chances of having the recipient open an attachment to a seemingly legitimate
message sent by a known acquaintance \cite{Oldfield-2001}.  Sometimes the
malware's originators use similar techniques to seed the attack. A more recent
example is `Operation Aurora', a sophisticated attack originating in China
against dozens of US companies during the first half of 2009, where the attack
was initiated via links spread through a popular Korean Instant Messaging
application \cite{AFP-2007-1}. Nevertheless, the current discussion focuses
more on the second context --- in which the human network structure itself is the
goal of the attack.

When discussing the goal of learning a network's
structure, it is important to distinguish between the ``technical'' topology of a
digital network and the actual topology of the human network that communicates on
top of it --- which is what we are actually interested in. Technically, every phone
or computer can reach nearly any other on the planet, but in practice it will
only contact a small subset, based on the context of its user. Many existing
network attacks gather information on the digital network topology, usually in
order to leverage the attack itself. Some attacks, for example, make use of an
email program's address book or a mobile phone's contact list to spread further.
In the context of Stealing Reality, this method is not as useful, since a
majority of peers would not be contacted on a routine basis. There is a great
deal of information in the patterns of communication exercised by the user with
his peers. These patterns are affected by many factors of relationship and
context, and could be used in reverse --- to infer the relationship and context. In
addition the communication patterns, combined with other behavioral data that can
be harvested from mobile devices, could serve to teach a great deal of
information about the user himself --- their age, their occupation and role,
their personality, and a great deal more. This type of information could be
summarized as a ``rich identity'' profile of a person
\cite{RealityMining}, which is much more informative than direct
demographic information which is currently used to profile users, and could be
very valuable to advertisers and spammers, for example.

Expanding from an individual's egotistical network, the social network as a whole
has intricate relationships and topologies among cliques and sub-groups, which
may be both overlapping as well as residing in multiple hierarchies. This is
complicated even more by issues of like trust or influence. The fact that three
people know each other does not necessarily mean that information received by one
will propagate in the same format to the two peers, if at all. Computational
social science has shown that many of these aspects of a social network could be
learned and extracted from communication patterns \cite{RealityMining}.

In this paper we discuss the ability to steal vital pieces of information
concerning networks and their users, by a non-aggressive (and hence --- harder to
detect) malware agent. We analyze this threat and build a mathematical model
capable of predicting the optimal attack strategy against various networks. Using
data from real-world mobile networks we show that indeed, in many cases a
``stealth attack'' (one that is hard to detect, however, and steals private
information at a slow pace) can result in the maximal amount of overall knowledge
captured by the operator of this attack.  This attack strategy also makes sense
when compared to the natural human social interaction and communication patterns,
as we discuss in our concluding section. The rest of the
paper is organized as follows: Sections \ref{sec.motivation} and
\ref{sec.dangers} expand on the motivation behind reality stealing attacks and
their dangers. Section \ref{sec.threat} describes the threat model and its
analysis, while Section \ref{sec.results} presents our preliminary empirical
results. Concluding remarks are given in Section \ref{sec.conc}.

\begin{figure*}[htb]
   \centering \includegraphics[width=\textwidth]{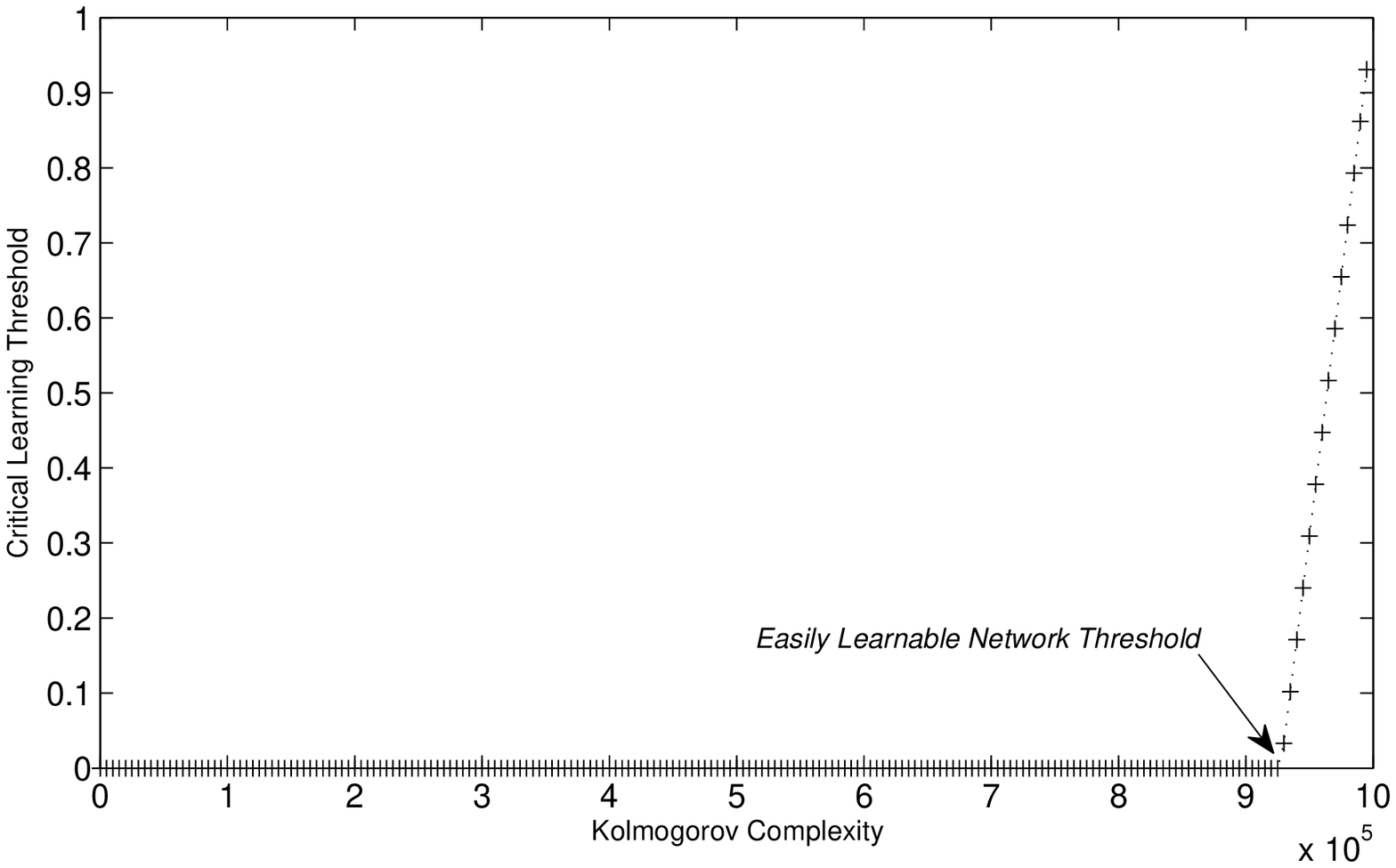}
   \caption{An illustration of the \emph{easily learnable network} notion. The
   graph depicts the critical learning threshold $\widehat{\Lambda_{E}}$ for
   networks of $1,000,000$ nodes, as a function of increasing values of the
   Kolmogorov complexity. Notice how networks for which $K_{E} < \max\left\{0 ,
   |E| - \frac{|E|}{\ln(|E|)}\right\}$ are easily learnable, while more complex
   networks require significantly larger amounts of information in order to be
   able to accelerate the network learning process. }
   \label{fig.res3}
\end{figure*}

\begin{figure*}[htb]
   \centering \includegraphics[width=\textwidth]{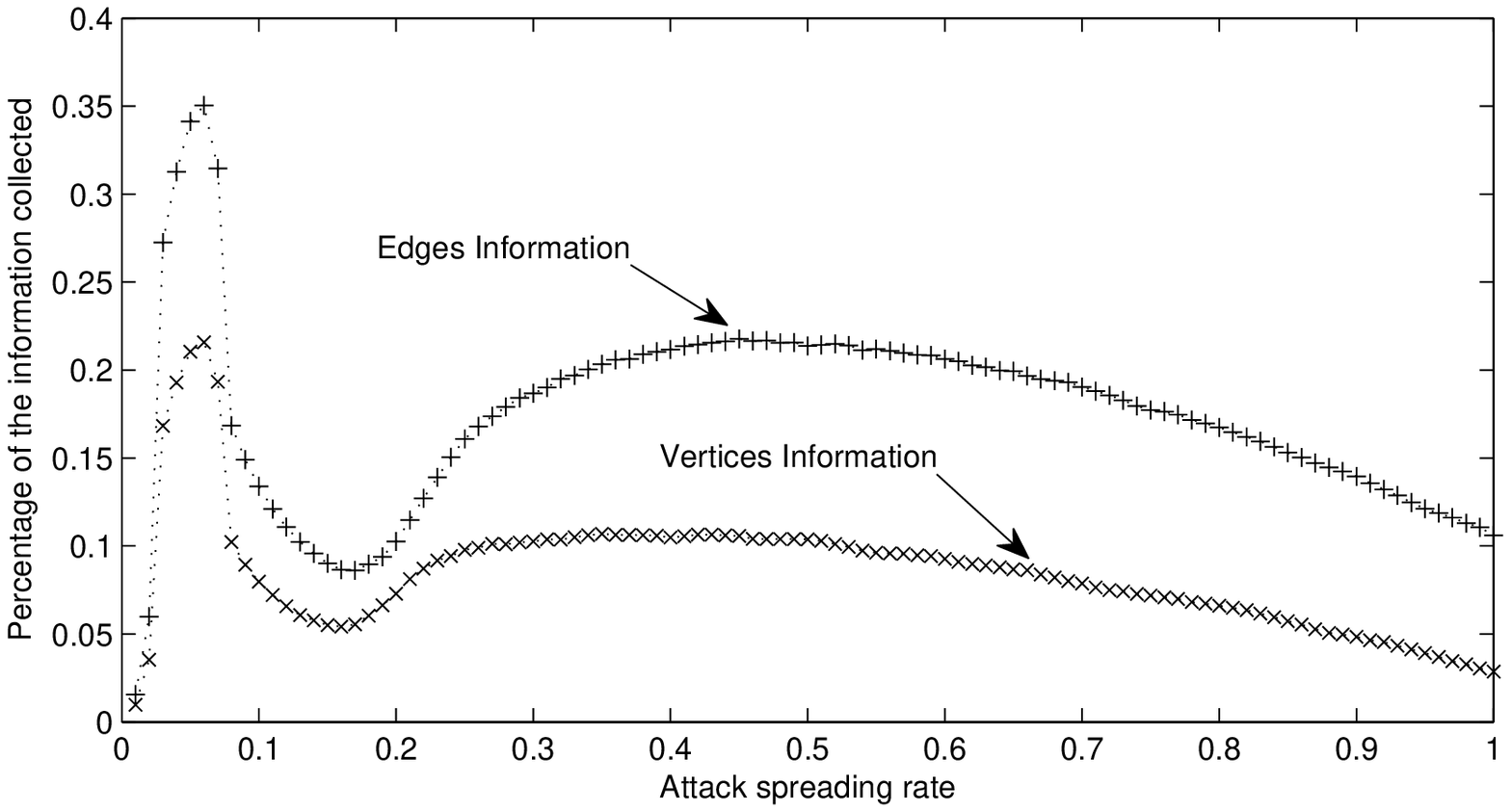}
   \caption{An observational study of the overall amount of data that can be
   captured as a function of $\rho$ --- the attack's aggressiveness. Notice the
   local maximum around $\rho = 0.5$ that is outperformed by the global maximum
   at $\rho = 0.04$. }
   \label{fig.res1}
\end{figure*}

\section{Motivation for Stealing Reality}\label{sec.motivation}

Many commercial entities have realized the value of information derived from
communication and other behavioral data for a great deal of applications, like
marketing campaigns, customer retention, security screening, recommender systems,
etc. There is no reason to think that developers of malicious applications will
not implement the same methods and algorithms into future malware, or that they
have not already started doing so. 

There already exist secondary markets for resale of this type of information,
such as \texttt{infochimps.com}, or black market sites and chat-rooms for resale
of stolen identity information and other illegal data sets
\cite{Herley_nobodysells}. It is reasonable to assume that a social hub's email
address would worth more to an advertiser than an edge node. It is also reasonable to
assume that a person meeting the profile of a student might be priced differently
than that of a corporate executive or a homemaker. There are already companies
operating in the legal grey area, which engage in the collection of email and
demographic information with the intention of selling it
\cite{flexo-email-addresses}. Why work hard when one can set loose automatic
agents that would collect the same if not better quality information? Wang et al.
 predict that once the market share of any specific mobile operating system
 reaches a computable phase transition point, viruses could pose a serious
 threat to mobile communications
 \cite{wang2009understanding}.

One might also imagine companies performing this types of attacks on a
competitor's customers (to figure out which customers to try and recruit), or
even operations performed by one country on another. Finally, the results of an
SR attack might be later used for selecting the best targets for future attacks
or configuring the `social engineering' components of other attacks.

\section{Why Stealing Reality Attacks Are So Dangerous}\label{sec.dangers}

One of the biggest risks of real world social network information being stolen is
that this type is very static, especially when compared to traditional targets of
malicious attacks. Data network topologies and identifiers could be replaced with
the press of a button. The same goes for passwords, usernames, or credit cards.
An infected computer could be wiped and re-installed. An online email, instant
messenger, or social networking account could be easily replaced with a similar
one, and the users' contacts can be quickly warned of the original account's
breach.

However, it is much harder to change one's network of real world,
person-to-person relationships, friendships, or family ties. The victim of a
``behavioral pattern'' theft cannot easily change her behavior and life patterns.
Plus this type of information, once out, would be very hard to contain. In addition,
once the information has been extracted in digital form, it would be quite hard
if not impossible to make sure that all copies have been deleted.

There are many stories in recent years of ``reality'' information being stolen
and irreversibly be put in the open. In 2008, real life identity information of
millions of Korean citizens was stolen in a series of malicious attacks and
posted for sale \cite{AFP-2007-1}. In 2007, Israel Ministry of Interior's
database with information on all of the country's citizens was leaked and posted
on the Web \cite{Jeffay-2009}.  Just these days, a court sill has to rule whether
the database of bankrupt gay dating site for teenagers will be sold to raise
money for repaying its creditors. The site includes personal information of over
a million teenage boys \cite{Emery-2010}. In all of these cases, once the
information is out, there is no way back, and the damage is felt for a long time
thereafter. In a recent Wall Street Journal interview, Google CEO Eric
Schmidt referred to the possibility that people in the future might choose to
legally change their name in order to detach themselves from embarrassing
``reality'' information publically exposed in social networking sites
\cite{ShmidtNameChange}. Speculative as this might be, it demonstrates the
sensitivity and challenges in recovering from leakage of real-life information,
whether by youthful carelessnes or by malicious extraction through an attack.

For this reason, Stealing Reality attacks are much more dangerous than
traditional malware attacks. The difference between SR attacks vs. more
traditional forms of attacks should be treated with the same graveness as
nonconventional weapons compared to conventional ones. The remainder of this
document presents our initial analysis and results regarding the form that an SR
attack might take, in contrast to the characteristics of conventional malware
attacks.

\section{Threat Model}\label{sec.threat}

In this section we describe and analyze the threat model. First, we define the
attacker's goals in the terms of our model, and develop a quantitative measure
for assessing the progress in achieving these goals. Then, we present an
analytical model to predict the success rate of various attacks. Finally, we
provide an assessment for the best strategies for devising such an attack. We
demonstrate both based on analytical models as well as using real mobile network
data, that in many cases the best attack strategy is counter intuitively a
``low-aggressiveness attack''. Besides yielding the best outcome for the
attacker, such an attack may also deceive existing monitoring tool, due to its
low traffic volumes and the fact that it imitates natural end-user communication
patterns (or even ``piggibacks'' on actual messages).

\subsection{Network Model}\label{sec.sub.network}

We shall model the network as an undirected graph $G(V,E)$. The difficulty of
learning the relevant information of the network's nodes and edges may be
different for different nodes and for different edges. In general, we denote the
probability that vertex $u$ was successfully ``learned'' or ``acquired'' by an
attacking agent that was installed on $u$ at time 0 as $p_{V}(u,t)$. Similarly,
we shall denote the probability that an edge $e(u,v)$ was successfully learned at
time $t$ by an agent installed on it at time 0 as  $p_{E}(u,t)$. We shall denote
the presence of an attacking agent on a vertex $u$ at time $t$ by the following
Boolean indicator:

\[
I_{u}(t) = 1 \emph{ iff $u$ is infected at time $t$}
\]

Similarly, we shall denote the presence of an attacking agent on an edge $e(u,v)$
at time $t$ as:

\[
I_{e}(t) = 1 \emph{ iff either $u$ or $v$ or both are infected at time $t$}
\]

For each vertex $u$ and edge $e$, let the times $T_{u}$ and $T_{e}$ denote their
initial time of infection.

\subsection{Attacker's Goal: Stealing Reality}\label{sec.sub.goal}

As information about the network itself has become a worthy cause for an attack,
the attacker's motivation is stealing as much properties related to the network's
social topology as possible. The percentage of vertices-related information
acquired at time $t$ is therefore:

\[
\Lambda_{V}(t) = \frac{1}{|V|} \sum_{u \in V} I_{u}(t) \cdot p_{V}(u,t-T_{u})
\]

Similarly, the percentage of edges-related information acquired at time $t$ is :

\[
\Lambda_{E}(t) = \frac{1}{|E|} \sum_{e \in E} I_{e}(t) \cdot p_{E}(e,t-T_{e})
\]

As an extension in the spirit of \emph{Metcalfe's} \cite{metcalfe} and
\emph{Reed's Law} \cite{reed-packlaw-2001}, a strong value emerges from learning the
``social principles'' behind a network. Understanding essence behind the implied
social network is more valuable (and also requires much more information in order
to learn) as the information it encapsulates is greater.  For example, let us
imagine the following two mobile social networks:
\begin{enumerate}
  \item For every two users $u_{i}$, $u_{j}$, the users are connected if and only
  if they joined the network on the same month.
  \item For every two users $u_{i}$, $u_{j}$, the users are connected in
  probability $p=\frac{1}{2}$.
\end{enumerate}

\begin{figure*}[htb]
   \centering
   \includegraphics[width=\textwidth]{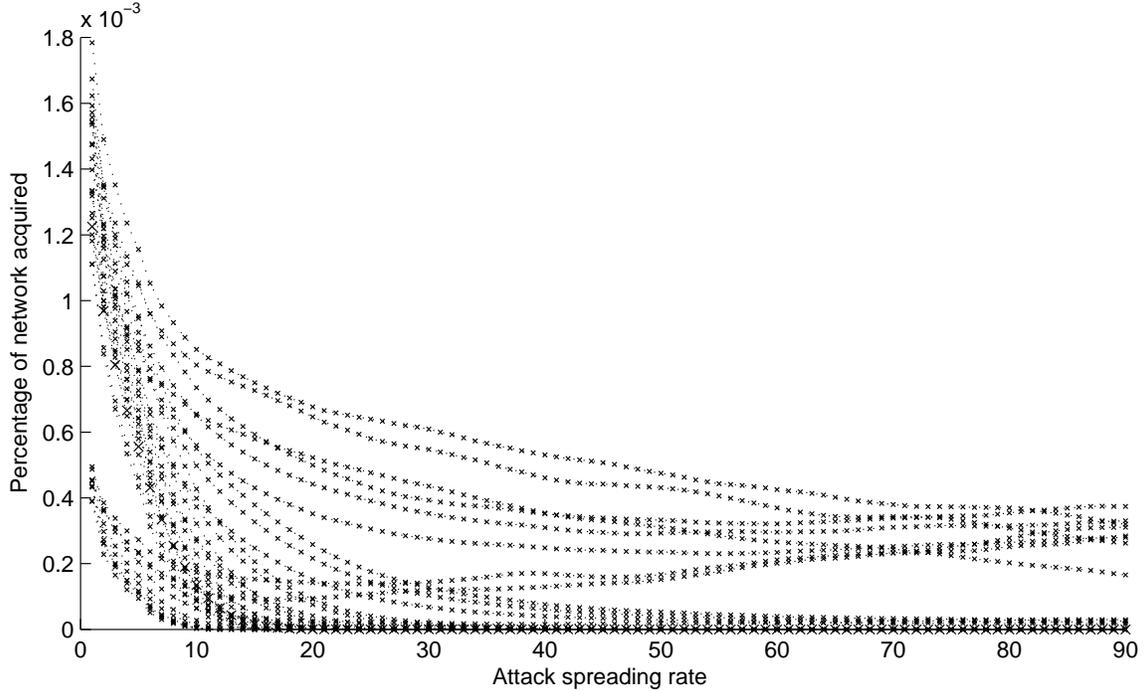}   
   \caption{An extensive study of a real-life mobile network of $7,706$ nodes and
   $17,404$ edges. Each graph presents the performance of a Stealing Reality
   attack for a specific different set of values of $\alpha, \beta, \sigma, M,
   r_{i}$. The performance is measured as the percentage of information acquired,
   as a function of the infection rate $\rho$. The scenarios that are presented
   in this figure demonstrate a global optimum of the attack performance for very
   low values of $\rho$. In other words, for many different scenarios it is best
   to use a very non-aggressive attack, which would result in maximizing the
   amount of network information obtained. Values of $\alpha$ and $\beta$ which
   had demonstrated this behavior were between $10$ and $500$. Values of $r_{i}$
   were between $0.1$ and $100$, whereas the values of $\sigma$ were between
   $0.1$ and $12$. The values of $M$ were between $0.1$ and $30$. It is
   interesting to mention that for high values of $\alpha$ and $\beta$, low
   values of $M$ did display this phenomenon while high values of $M$ did not.
   }
   \label{fig.res4}
\end{figure*}

It is easy to see that given a relatively small subset of network $1$, the logic
behind its social network can be discovered quite easily. Once this logic is
discovered, the rest of the network can automatically be generated (as edges are
added exactly for pairs of users who joined the network at the same month).
Specifically, for every value of $\epsilon$ we can calculate a relatively small
number of queries that we should ask in order to be able to restore the complete
network with mistake probability of $1-\epsilon$. However, for network $2$ the
situation is much different, as the only strategy for accurately obtaining the
network is actually discovering all the edges it comprised of.

Let us denote by $K_{E}$ the \emph{Kolmogorov Complexity}
\cite{KolmogorovComplexity} of the network, namely --- the minimal number of bit
required in order to ``code'' the network in such a way that it could later be
completely restored. As the number of vertices $|V|$ is assumed to be known, the
essence of the network is coded in its edges. Dividing the number of edges
learned $|E| \Lambda_{E}(t)$ by the number of ``redundant edges'' $|E| - K_{E}$
yields the amount of information learned at time $t$. Following a similar logic
of \emph{Reed's Law} we shall evaluate the benefit of the learning process
proportionally to the number of combinations that can be composed from the
information learned. Normalizing it by the number of edges, we shall receive the
following measurement for the social essence learned:

\[
\Lambda_{S}(t) = 2^{\frac{|E| \Lambda_{E}(t) - |E|}{|E| - K_{E}}}
\]

The attacker in interested in maximizing the values of $\Lambda_{V}(t)$,
$\Lambda_{E}(t)$ and $\Lambda_{S}(t)$. The evolution of the $\Lambda_{S}$, the
social essence of the network, as a function of the ``complexity hardness'' of
the network is illustrated in Figure \ref{fig.res2}.

\subsection{Attack Analysis}\label{sec.sub.analysis}

We assume that the learning process of vertices and edges follows the well-known
\emph{Gompertz} function, namely:

\[
\forall e_{t} \in E \quad , \quad p_{E}(e,t) = e^{- \alpha e^{-r_{i}(t)}}
\]

\[
\forall u_{t} \in V \quad , \quad p_{V}(u,t) = e^{- \beta e^{-r_{i}(t)}}
\]
with $\alpha$ and $\beta$ representing the efficiency of the learning mechanism
used by the attacker, as well as the amount of information that is immediately
obtained upon installation. $r_{i}$ denotes the learning rate of each edge \
vertex,  determined by the activity level of the edge \ vertex (namely ---
accumulation of new information). Variable $r_{i}$ is also used for normalizing $\alpha$
and $\beta$. Hence, the attack success rates can now be written as follows:

\[
\Lambda_{V}(t) = \frac{1}{|V|} \sum_{u_{i} \in V} I_{u_{i}}(t) \cdot e^{-\beta
e^{-r_{i}(t - T_{u_{i}})}}
\]

\[
\Lambda_{E}(t) = \frac{1}{|E|} \sum_{e_{i} \in E} I_{e_{i}}(t) \cdot e^{-\alpha
e^{-r_{i}(t - T_{e_{i}})}}
\]

Attacking agents spread through movements on network edges. Too aggressive
infection is more likely to be detected, causing the accumulation of information
concerning the network to be blocked altogether. On the other hand, attack agents
that spread too slowly may evade detection for a long period of time, however,
the amount of data they gather would still be very limited. In order to predict
the detection probability of attacking agents at time t we shall use
\emph{Richard's Curve}, for as follows :

\[
p_{detect}(t) = \frac{1}{\left(  1 + e^{-\rho(t-M)} \right)^{\frac{1}{\rho}\sigma}}
\]

where $\rho$ is the probability that an agent would copy itself to a neighboring
vertex at each time step, $\sigma$ is a normalizing constant for the detection
mechanism, and $M$ denotes the normalizing constant for the system's initial
state. Let $N_{t}$ denote the number of infected vertices at time $t$. Assuming
that vertices infection by their infected neighbors is a random process, the
number of infected vertices vertex $u$ would have at time $t$ is~:

\[
N_{t} \cdot \frac{deg(u)}{|V|}
\]

The probability that vertex $u$ would be attacked at time $t$ equals therefore at least:

\[
p_{attack}(u,t) = 1 - e^{-N_{t} \cdot \rho \cdot \frac{deg(u)}{\sum_{v \in V} deg(v)}}
\]

and the expected number of infected vertices is :

\[
N_{t + \Delta t} = |V| - \sum_{v in V}  \prod_{i=0}^{t} (1 - p_{attack}(v,i))
\]

The number of infected nodes therefore grows as :

\[
N_{t + \Delta t} = |V| - \sum_{v in V}  \prod_{i=0}^{t} e^{-N_{i} \cdot \rho \frac{deg(v)}{2|E|}}
\]

From $N_{t}$ we can now derive the distribution of the Boolean infection indicators :

\[
p[I_{u}(t) = 1] = \frac{N_{t}}{|V|}
\]

\[
p[I_{e}(t) = 1] = 2 \frac{N_{t}}{|V|} - \frac{N_{t}^{2}}{|V|^{2}}
\]

And the attack probability can now be given as follows :

\[
p_{attack}(u,t+\Delta t) = \]
\[ 1 - e^{\rho \frac{deg(u)}{2|E|} \left( -|V| + \sum_{v \in V}
\prod_{i=0}^{t}(1-p_{attack}(v,i))\right)}
\]

This expression can now be used for calculating the distribution of initial
infection times of vertices and edges. Note that information is gathered faster
as infection rate $\rho$ increases. However, so does the detection probability.
The optimum can therefore be derived by calculating the expectance of the total
amount of information obtained (in which the only free parameter is $\rho$) :

\[
\Lambda_{E} = \int_{0}^{\infty} \left( \frac{\partial \Lambda_{E}(t)}{\partial t}
\cdot (1 - p_{detect}(t)) \right)dt
\]

\[
\Lambda_{V} = \int_{0}^{\infty} \left( \frac{\partial \Lambda_{V}(t)}{\partial t}
\cdot (1 - p_{detect}(t)) \right)dt
\]

\subsection{Obtaining the Social Essence of a Network}\label{sec.sub.essence}

Recalling the expression that represents the progress of learning the ``social
essence'' of a network, we can see that initially each new edge contributes
$O(1)$ information, and the overall amount of information is therefore kept
proportional to $O(\frac{1}{|E|})$. As the learning progresses and the logic
principles behind the network's structure start to unveil, the amount of
information gathered from new edges becomes greater than their linear value. At
this point, the overall amount of information becomes greater than
$O(\frac{1}{|E|})$, and the benefit of acquiring the social structure of the
network starts to accelerate. Formally, we can see that this phase is reached
when:

\[
\Lambda_{E}(t) > O\left(1 - \frac{|E| - K_{E}}{|E|} \ln(|E|)\right)
\]

Let us denote by $\widehat{\Lambda_{E}}$ the \emph{Critical Learning Threshold},
above which the learning process of the networks accelerates, as described above
(having each new learned edge contributing an increasingly growing amount of
information concerning the network's structure), to be defined as follows:

\[
\widehat{\Lambda_{E}} \triangleq 1 - \frac{|E| - K_{E}}{|E|} \ln(|E|)
\]

Consequently, in order to provide as strong protection for the network as
possible, we should make sure that for every value of $t$:

\[ \sum_{e_{i} \in E} I_{e_{i}}(t) \cdot e^{-\alpha e^{-r_{i}(t - T_{e_{i}})}} <
|E| - \left(|E| - K_{E} \right) \ln(|E|) \]

Alternatively, the attack would prevail when there exist a time $t$ for which the
above no longer holds.

Notice that as pointed out above, ``weaker'' networks (namely, networks of low
Kolmogorov complexity) are easy to learn using a limited amount of information.
Generalizing this notion, the following question can be asked : How ``simple'' must a network be, in order for it to be ``easily learnable''
(namely, presenting an superlinear learning speed, starting from the first edges
learned)?

It can be seen that in order for a network to be easily learnable, its critical
learning threshold $\widehat{\Lambda_{E}}$ must equal $O(1)$. Namely, the
network's Kolmogorov complexity must satisfy: \[ 1 - \frac{|E| - K_{E}}{|E|}
\ln(|E|) < O(1) \]

We must obtain the following criterion for \emph{easily learnable networks}: \[
K_{E} < |E| - \frac{|E|}{\ln(|E|)} \] The notion of an \emph{easily learnable
network} is illustrated in Figure \ref{fig.res3}, presenting the critical
learning threshold $\widehat{\Lambda_{E}}$ for networks of $1,000,000$ nodes, as
a function of the network's Kolmogorov complexity.

\section{Experimental Results}\label{sec.results}

We have evaluated our model on aggregated call logs derived from a real mobile phone network, comprised
of approximately $200,000$ nodes and $800,000$ edges. These tests have clearly
shown that in many cases, an ``aggressive attack'' achieves inferior results
compared to more subtle attacks. Furthermore, although sometimes the optimal
value for the infection rate revolves around $50\%$, there are scenarios in which
there is a local maximum around this value, with a global maximum around $4\%$.
Figure \ref{fig.res1} demonstrates the attack efficiency (namely, the maximal
amount of network information acquired) as a function of its ``aggressiveness''
(i.e. the attack's infection rate). A global optimum both for the vertices
information as well as for the edges information is achieved around $4\%$, with a
local optimum around $52\%$.

A more extensive simulation research was conducted for an arbitrary sub-network
of this mobile network, containing $7,706$ edges and $17,404$ edges. In this
research we have extensively studied the success of a Stealing Reality attack
using numerous different sets of values (i.e. $\alpha$, $\beta$, $r_{i}$,
$\sigma$ and $M$). Although the actual percentage of stolen information had
varied significantly between the various sets, many of them had displayed the
same interesting phenomenon --- a global optimum for the performance of the
attack, located around a very low value of $\rho$. Some of these scenarios are
presented in Figure \ref{fig.res4}.

\section{Conclusions}\label{sec.conc}

In this paper we discussed the threat of malware targeted at extracting
information about the relationships in a real-world social network as well as
characteristic information about the individuals in the network, which we name
``Stealing Reality''. We present Stealing Reality (SR), explain why it differs
from traditional types of network attacks, and discuss why its impact is
significantly more dangerous than that of other attacks. We also present our
initial analysis and results regarding the form that an SR attack might take. We
have evaluated our model on data derived from a real mobile network. Our results
clearly show that an ``aggressive attack'' achieves inferior results compared to
more subtle attacks. This attack strategy also makes sense when comparing it to
natural human social interaction and communication patterns. The rate of human
communication and evolution of relationship is very slow compared to traditional
malware attack message rates. A Stealing Reality type of attack, which is
targeted at learning the social communication patterns, could ``piggyback'' on
the user generated messages, or imitate their natural patterns, thus not drawing
attention to itself while still acheiving its target goals.

\bibliography{Altshuler}
\end{document}